\documentclass{optica-article}

\journal{opticajournal} 

\articletype{Research Article}

\usepackage{stfloats}
\usepackage{lineno}
\usepackage{amsmath}
\usepackage{float}

\begin{document}

\title{Cross-attention learning enables real-time nonuniform rotational distortion correction in OCT}

\author{\textcolor{purple}{Haoran Zhang,\authormark{1} Jianlong Yang,\authormark{1,*} Jingqian Zhang,\authormark{1} Shiqing Zhao,\authormark{1} and Aili Zhang\authormark{1}}}

\address{\authormark{1}School of Biomedical Engineering, Shanghai Jiao Tong University, Shanghai, China\\
\authormark{*}Corresponding Author: Jianlong Yang (jyangoptics@gmail.com)}


\begin{abstract*} 
Nonuniform rotational distortion (NURD) correction is vital for endoscopic optical coherence tomography (OCT) imaging and its functional extensions, such as angiography and elastography. Current NURD correction methods require time-consuming feature tracking/registration or cross-correlation calculations and thus sacrifice temporal resolution. Here we propose a cross-attention learning method for the NURD correction in OCT. Our method is inspired by the recent success of the self-attention mechanism in natural language processing and computer vision. By leveraging its ability to model long-range dependencies, we can directly obtain the spatial correlation between OCT A-lines at any distance, thus accelerating the NURD correction. We develop an end-to-end stacked cross-attention network and design three types of optimization constraints. We compare our method with two traditional feature-based methods and a CNN-based method on two publicly-available endoscopic OCT datasets. We further verify the NURD correction performance of our method on 3D stent reconstruction using a home-built endoscopic OCT system. Our method achieves a $\sim3\times$ speedup to real time ($26\pm 3$ fps), and superior correction performance.

\end{abstract*}

\section{Introduction}
\indent Optical coherence tomography (OCT) \cite{huang1991optical} uses temporal coherence gating to resolve depth information in micrometer scale. It enables non-invasive tomographic imaging of biological tissues with near-cellular spatial resolution and high sensitivity \cite{drexler2014optical}. Nowadays OCT has become a routine diagnostic instrument in ophthalmology \cite{adhi2013optical}. Through a fiber-optic endoscopic probe, its application is expanding to other medical fields for \textit{in situ} label-free biopsy, such as cardiovascular, respiratory, gastrointestinal, and cervix sites \cite{swanson2017ecosystem,zagaynova2008endoscopic,bouma2017intravascular}.\\
\indent For such applications, an endoscopic probe with point-by-point scanning capability is usually required. Typically, the scanning is controlled externally and implemented mechanically to achieve axial movement and circumferential rotation of the probe (referred to as proximal scanning). In recent years, with the development of technologies such as MEMS and piezoelectric devices, point-by-point scanning can be achieved by shifting the beam at the output end of the probe (referred to as distal scanning). However, distal scanning is currently rarely used clinically due to its significantly higher cost and larger size of the probe compared to the proximal scanning \cite{gora2017endoscopic}.\\
\indent Due to the irregularities in the shape of vessels and other lumen structures, friction, and torque transmission losses, the rotation of the proximal scanning probe becomes non-uniform, resulting in distortion of the intracanal OCT images, known as non-uniform rotational distortion (NURD) \cite{araki2022optical}. NURD can introduce errors in the morphological representation of tissues and make it difficult to perform functional imaging of tissue, such as elasticity, birefringence, angiography, and treatment processes \cite{ahsen2014correction,wang2015heartbeat,lo2019balloon}. Effective NURD correction is demanded to deal with such problems. For many application scenarios of OCT that require real-time operation or fast evaluation, such as surgical robot navigation, online monitoring of treatment, and \textit{in situ} diagnosis, the time cost of the NURD correction should be considered. \\
\indent Existing methods for the NURD correction are primarily based on feature tracking/registration and dynamic programming \cite{lo2019balloon,cao2023improved,van2008azimuthal,qi2021automatic}. William \textit{et al.} used the speeded-up robust feature (SURF) operator to extract feature points in OCT B-frames and then tracked them across adjacent frames for A-line alignment \cite{lo2019balloon}. Cao \textit{et al.} proposed an improved feature extraction algorithm and put it into coarse and fine registration process \cite{cao2023improved}. These methods rely on extracting a large number of feature points to improve the correction accuracy. Therefore, there is a trade-off between the time cost of feature extraction and accuracy. Soest \textit{et al.} utilized the dynamic programming method to find a continuous path through a spatial cross-correlation matrix that measures the region similarity between adjacent frames \cite{van2008azimuthal}. However, the construction of the cross-correlation matrix is time-consuming. Qi \textit{et al.} used a graph-based dynamic programming algorithm to find an optimal path that represents the initial rotation angle error drifting along the pull-back direction \cite{qi2021automatic}, which significantly speeds up the processing, but the A-line level distortion is neglected.\\
\indent Other methods utilized hardware and prior knowledge specific to the endoscopic probe or imaging target \cite{ahsen2014correction,miao2021graph}, thus lacking generality. Abouei \textit{et al.} presented a motion artifact correction method based on azimuthal en face image registration \cite{abouei2018correction}. However, this method needs to collect the complete image sequence first and thus cannot correct the distortion in real time. Uribe-Patarroyo and Bouma developed a method based on speckle decorrelation \cite{uribe2015rotational}, which could perform NURD correction in real-time, but the decorrelation is vulnerable to the variation of environment, such as motion and temperature \cite{guo2019intraoperative}.\\
\indent Recently, Liao \textit{et al.} proposed a convolutional neural network (CNN)-based learning method for the NURD correction \cite{liao2022distortion}. They developed a new A-line level shifting error vector estimation network to extract the optimal path from a spatial correlation matrix. Another CNN branch was introduced to suppress the accumulative error. Their method outperforms previous ones on correction performance and achieved a processing rate of around 7 fps (frame per second). However, CNNs have limitations in modeling long-range dependencies due to the constraints of local receptive fields and fixed convolutional kernel sizes, thus requiring pre-build a spatial correlation matrix as network input, which affects their capability to scale up the processing efficiency.\\
\indent In this work, we propose a cross-attention learning method to address the limitations of existing NURD correction methods above. Our method is inspired by the recent success of the self-attention mechanism \cite{vaswani2017attention} in natural language processing (NLP) and computer vision (CV), which has played a crucial role in the development of cutting-edge tools like ChatGPT \cite{stiennon2020learning}. Our key finding here is that the self-attention mechanism enables the direct establishment of global spatial correlations within OCT A-line sequences, without the necessity of correlation calculation in advance. Because the self-attention mechanism is used between different A-lines, we refer to it here as cross-attention. To achieve a high correction efficiency, we develop an end-to-end stacked cross-attention network and design three types of optimization constraints.
\section{Methods and materials}
\subsection{Overall framework}
\begin{figure}[htb]
    \centering
    \centering\includegraphics[width=10cm]{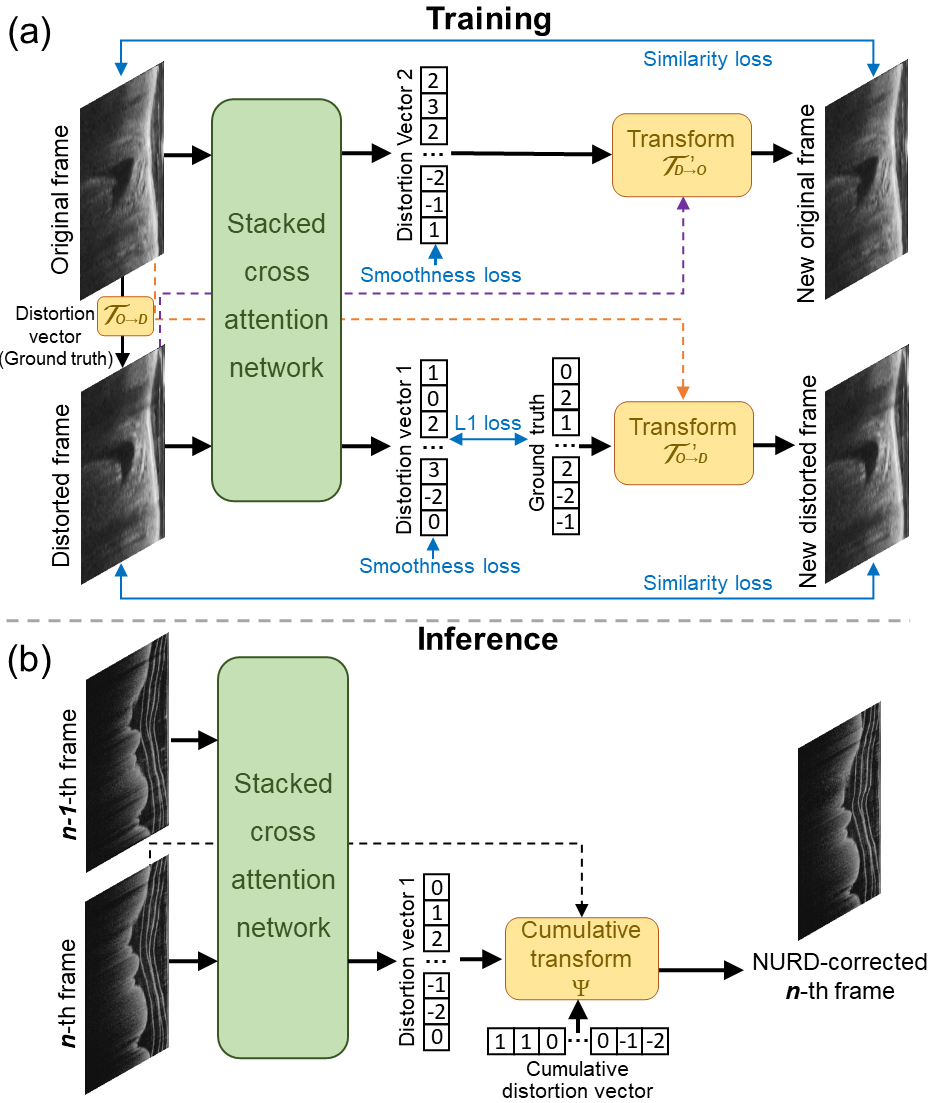}
    \caption{Overall framework of our proposed method. (a) and (b) illustrate its training and inference phases, respectively.}
    \label{fig:img1}
\end{figure}
\begin{figure}[htb]
    \centering
    \centering\includegraphics[width=13cm]{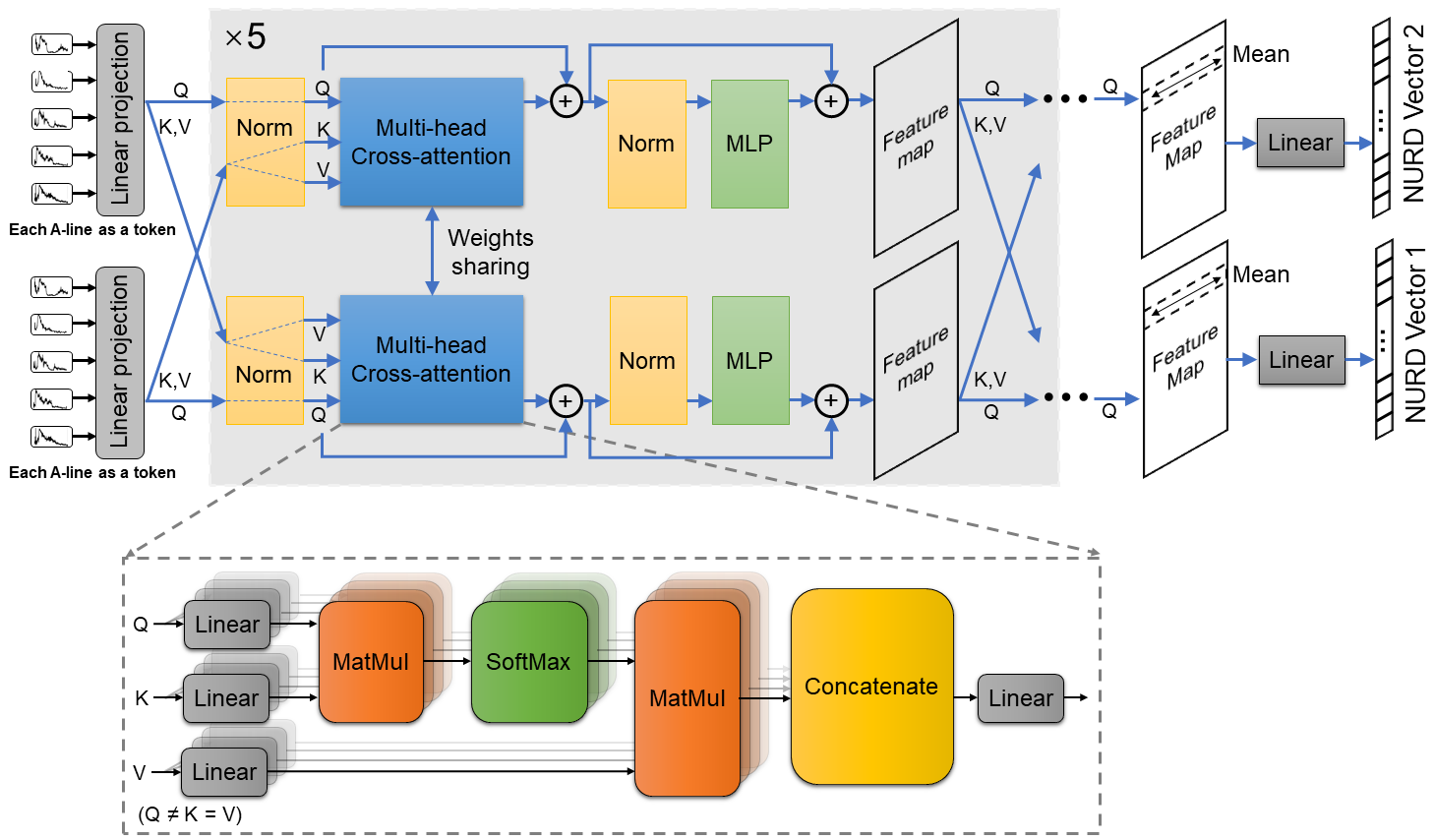}
    \caption{Illustration of the stacked cross-attention network. The upper panel is the overall architecture and the lower dashed box is the details of the multi-head cross-attention module.}
    \label{fig:img2}
\end{figure}

\indent Figure~\ref{fig:img1} illustrates the overall framework of our proposed method. (a) and (b) illustrate its training and inference phases, respectively. We use a self-supervised generative learning approach for training, \textit{i.e.}, by distorting the original B-scans and then using the network to predict their distortions. Specifically, in Fig.~\ref{fig:img2}, we use a distortion vector, which serves as the ground truth (GT) of the A-line shifts due to the NURD, to do the transform ${\rm{{\cal T}}}_{O\rightarrow D}$ from the original frame to the distorted frame (the generation of the GT distortion vectors follows the method described in Section 3.2.1 of \cite{liao2022distortion}). These two frames are then fed into the stacked cross-attention network, which is employed to correct the NURD. It predicts two distortion vectors: the first one is the distortion applied on the original frame to form the distorted frame; the second one is the distortion applied on the distorted frame to form the original frame. This bi-directional design is inspired by the notion of cycle consistency in generative learning \cite{zhu2017unpaired}. Using these predicted vectors, we can apply the transform ${\rm{{\cal T}}}^{\prime}_{O\rightarrow D}$ and ${\rm{{\cal T}}}^{\prime}_{D\rightarrow O}$ to the original and distorted frames, and form the new distorted and original frames, respectively. We use three types of optimization constraints in the training: (1) mean absolute error loss (L1 loss) $\mathcal{L}_{l1}$ between the distortion vector 1 and the GT vector, (2) smoothness loss $\mathcal{L}_{sm}$ of the predicted distortion vectors, and (3) similarity loss $\mathcal{L}_{si}$ between the original/distorted frames and the new original/distorted frames at the A-line level. We list their functions below:
\begin{gather}
    \mathcal{L}_{l1} = \frac{1}{{{N}}}\sum\limits_{i = 1}^{{N}} {\left| {{{\hat d}_i} - {d_i}} \right|},\\
    \mathcal{L}_{sm} = \frac{1}{{{N} - 1}}\sum\limits_{i = 1}^{{N} - 1} {\left| {{{\hat d}_i} - {{\hat d}_{i + 1}}} \right|},\\
    \mathcal{L}_{si} = \frac{1}{N}\sum\limits_{i = 1}^N {\left| {\frac{1}{M}\sum\limits_{j = 1}^M {{{\hat p}_{i,j}}}  - \frac{1}{M}\sum\limits_{j = 1}^M {{p_{i,j}}} } \right|},
\end{gather}
where ${\hat d}_i$ and $d_i$ are the elements of the predicted distortion vector and ground truth, respectively. $N$ is the length of the vector (also the number of A-lines in each frame). $M$ is the number of data points in each A-line. ${\hat p}_{i,j}$ and ${p}_{i,j}$ are the pixel value of data point $j$ in A-line $i$ from the predicted new frame and the corresponding input image, respectively. The smoothness loss and similarity loss are all adopted in the prediction of two distortion vectors, and L1 loss is only adopted in the prediction of distortion vector 1 because the GT vector of distorting original frame is known. The final loss of network is:
\begin{gather}
    \mathcal{L} = \mathcal{L}_1 + \mathcal{L}_{sm-1} + \mathcal{L}_{sm-2} + \mathcal{L}_{si-1} + \mathcal{L}_{si-2}.
\end{gather}
\indent In the inference phase, two successively acquired OCT B-scans (the raw $n-1$-th and $n$-th frames. $n$ refers to time points) are fed into the trained stacked cross-attention network. The output of this network is only the distortion vector 1, which is used to correct the NURD of the newest $n$-th frame. We generate the cumulative distortion vector from raw $n$-th frame to the initial $1$-th frame using the method described in \cite{liao2022distortion}. Specifically, the $n$-th frame is composed of $N$ A-lines $A_i^n$ ($i \in \left[ {1,N} \right]$). Due to the NURD occurrence in adjacent frames, $A_i^n$ mismatches its correct position which is supposed to be aligned to $A_j^{n-1}$. The position error $\varepsilon _i^n = j - i$ of A-line $A_j^{n}$ constitutes one element of distortion vector ${D^{n \sim n - 1}} = {\left[ {\varepsilon _1^n,...,\varepsilon _i^n,...,\varepsilon _N^n} \right]^T}$ (it can be integers only). Given predicted A-line level distortion vector ${{\hat D}^{n \sim n - 1}}$ between $n$-th and $n-1$-th frames and cumulative ${D^{n-1 \sim 1}}$ between $n-1$-th and initial $1$-th frames, the latest distortion vector ${D^{n \sim 1}}$ can be obtained by cumulative transform operation $\Psi$:\\
\begin{equation}
\label{eq3}
\begin{split}
D_i^{n \sim 1} &= {\Psi _{(i)}}(\hat D^{n \sim n - 1},D^{n - 1 \sim 1}) = \hat D_i^{n \sim n - 1} + D_j^{n - 1 \sim 1},\\
j &= \hat D_i^{n\ \sim n-1} + i.
\end{split}
\end{equation}
where we can cumulatively transform the $n$-th frame to the initial $1$-th frame in A-line level, and finally generate NURD-corrected $n$-th frame.

\subsection{Stacked cross-attention network}
\indent Figure~\ref{fig:img2} illustrates the stacked cross-attention network. (a) is the overall architecture and (b) is the details of the multi-head cross-attention module. Instead of 2D operations employed in CNNs, here we use each A-line (1D) of the OCT B-scans as a token. Then they are used to calculate the query (Q), key (K), and value (V) vectors in the self-attention mechanism \cite{vaswani2017attention}: 
\begin{align}
\mathbf{Q} &= \mathbf{X} \cdot \mathbf{W}_Q + \mathbf{b}_Q \\
\mathbf{K} &= \mathbf{X} \cdot \mathbf{W}_K + \mathbf{b}_K \\
\mathbf{V} &= \mathbf{X} \cdot \mathbf{W}_V + \mathbf{b}_V 
\end{align}
where \(\mathbf{X}\) (\(\mathbf{X} \in \mathbb{R}^{N \times M}\)) is the input tokens with \(N\) A-lines and \(M\) data points in each A-line. The linear projection is defined as:\(\mathbf{Q}, \mathbf{K}, \mathbf{V} \in \mathbb{R}^{N \times E}\) ($E$ is the embeding dimension, and $E>M$), and \(\mathbf{W}_Q, \mathbf{W}_K, \mathbf{W}_V\) are weight matrices, while \(\mathbf{b}_Q, \mathbf{b}_K, \mathbf{b}_V\) are bias terms. This linear projection step allows the model to capture different aspects of the input sequence. The query vectors \(\mathbf{Q}\) represent the current token and are responsible for computing attention weights. The key vectors \(\mathbf{K}\) capture the contextual information of each element, enabling the model to assess the relevance between different elements. The value vectors \(\mathbf{V}\) carry the actual content information associated with each token. Then these vectors are fed into 5 consecutive multi-head cross-attention blocks ($\times 5$). Each block includes a multi-head cross-attention module and a multi-layer perception (MLP) module \cite{vaswani2017attention}. In each block, we apply layer normalization (Norm) before each module and conduct residual connections. Finally, we perform averaging and linear operations to get the distortion vectors.\\
\indent The multi-head cross-attention module in the lower dashed box of Fig.~\ref{fig:img2} allows the model to attend to different parts of the input sequence and capture diverse dependencies, enhancing its representation and predictive capabilities. Given a sequence of input embeddings $\mathbf{X} = [\mathbf{x}_1, \mathbf{x}_2, \ldots, \mathbf{x}_n]$, the output is computed as follows:
\begin{equation}
    \text{MultiHead}(\mathbf{Q}, \mathbf{K}, \mathbf{V}) = \text{Concat}(\text{head}_1, \text{head}_2, \ldots, \text{head}_h) \mathbf{W}^O
\end{equation}
where $\text{head}_i = \text{Attention}(\mathbf{Q} \mathbf{W}^Q_i, \mathbf{K} \mathbf{W}^K_i, \mathbf{V} \mathbf{W}^V_i)$ represents the attention mechanism applied on the projected queries $\mathbf{Q} \mathbf{W}^Q_i$, keys $\mathbf{K} \mathbf{W}^K_i$, and values $\mathbf{V} \mathbf{W}^V_i$ of the $i$-th attention head. Here, $\mathbf{W}^Q_i$, $\mathbf{W}^K_i$, and $\mathbf{W}^V_i$ are learnable linear projection matrices specific to each attention head. The concatenated outputs are then linearly transformed by the matrix $\mathbf{W}^O$ to produce the final output.
\subsection{Datasets and implementations}
\begin{figure}[H]
    \centering
    \centering\includegraphics[width=13cm]{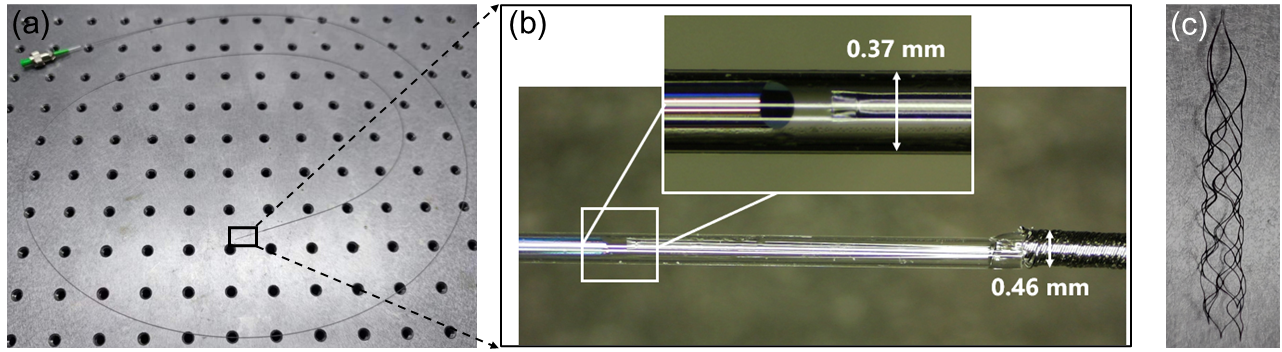}
    \caption{(a) Photograph of our assembled proximal scanning micro-probe used for endoscopic OCT imaging. (b) Enlarged view of the black box in (a). (c) Photograph of the intravascular stent used in OCT imaging.}
    \label{fig:img3}
\end{figure}

\indent We collect a total of 7,731 endoscopic OCT B-scans from publicly-available datasets \cite{wang2015heartbeat,uribe2015rotational,sun2012vivo,lee2011quantification,miao2021graph,li2012high,wang2013intravascular,yun2006comprehensive} to train our model. As mentioned above, we use these data to generate the GT distortion vectors using the method in \cite{liao2022distortion}. By applying these vectors to the B-scans, we create 20,000 original-distorted image pairs for the training. Because most of them are from clinical acquisition, the temporal and spatial characteristics of the distortion vectors are consistent with real application scenarios. We then use another two synthetic endoscopic datasets and two real publicly-available endoscopic datasets \cite{gora2013tethered,liao2021data} for evaluating our trained model. Note that we train our model in one go and evaluate it on external test datasets. Compared to the commonly used division of the same dataset into training and test sets, this approach can better demonstrate the accuracy and robustness of our approach and the generalization ability of the model. \\
\indent The synthetic endoscopic OCT sequences are employed because we cannot get the GT of NURD from real endoscopic OCT data. We follow the method described in \cite{liao2022distortion} to generate the synthetic sequences. Firstly, a motion (NURD)-free OCT sequence is created by repeating an OCT B-frame 500 times. Then we apply 499 random distortion vectors to all frames except for the first one. We employ a pig bronchus OCT B-scan \cite{li2012high} and a human nasopharynx OCT B-scan \cite{miao2021graph} (as shown in Fig.~\ref{fig:supplement2} below) to generate the two synthetic sequences for testing. The two real OCT sequences for testing include a gastrointestinal tract sequence (648 images) \cite{gora2013tethered} and a sponge surface sequence (240 images) \cite{liao2021data}.\\
\indent Besides, we further evaluate the NURD correction performance using our home-built endoscopic SD-OCT system. Our system has a central wavelength of $\sim840$ nm and a bandwidth of $\sim50$ nm, which corresponds to an axial resolution of $\sim5$ $\mu$m. Its A-line rate is 80 kHz. A homemade capillary tube-based fiber optic rotary joint \cite{kim2016lensless} driven by a commercial motor (34 rps rotation speed) is applied to perform circumferential scanning. As shown in Fig.~\ref{fig:img3}(a), an assembled proximal scanning micro-probe with 1.2 m length offers a lateral resolution of 25 $\mu$m and a working distance of 2 mm. The micro-probe with a transparent glass tube is 0.37 mm in diameter shown in the enlarged view of Fig.~\ref{fig:img3}(b). In the experiment, a 30 mm length intravascular stent with 4 mm diameter was used for imaging as shown in Fig.~\ref{fig:img3}(c).
\indent We implement the code of our proposed method using pyTorch. Our model is trained on a personal computer with an Nvidia 3090 GPU (24G onboard memory). We convert an endoscopic OCT B-scan into an input format where each frame consists of 1024 A-lines, and each A-line contains 512 data points. We employ the multi-head cross-attention with an embedding dimension of 1024 and 4 heads. We use the stochastic gradient descent (SGD) \cite{bottou2010large} optimizer with a learning rate of $5e-4$. We set a batch size of 24 and train our model for 200 epochs. The training time is about 33 hours. It should be noted that the model is trained once and for all.\\ 
\indent We use the mean absolute error (MAE) $\psi (n)$ to quantitatively evaluate the correction performance of the synthetic sequences:
\begin{equation}
\psi (n) = \frac{1}{N}\sum\limits_{i = 1}^N {\left| {\mathop {D_i^n}\limits^ \wedge   - D_i^n} \right|}
\end{equation}
where $\mathop {D_i^n}\limits^ \wedge$ and $D_i^n$ are the predicted and the GT shifts of \textit{i}-th Aline of distortion vectors within \textit{n}-th frame, respectively. For the real publicly-available sequences, because the GT of the distortion vector is unknown, we use the mean standard deviation (mean-STD) $\sigma \left( n \right)$ to quantitatively evaluate the correction performance, which was commonly adopted in previous NURD correction works \cite{van2008azimuthal,liao2022distortion}:
\begin{equation}
\sigma \left( n \right) = \frac{1}{{N \times M}}\sum\limits_{i = 1,j = 1}^{N \times M} {{{\tilde \sigma }_5}({p_{i,j}})}
\end{equation}
where ${\tilde \sigma }_5({p_{i,j}})$ is the mean-STD of pixel $p_{i,j}$ in adjacent 5 frames with $n$-th frame as the center. Precise correction can reduce the mean-STD to nearly 0, but it will never be exactly 0 due to variations in scanning locations and speckle/decorrelation noise.
\section{Results}
\subsection{Accuracy assessment of NURD correction}
\begin{figure}[H]
    \centering
    \centering\includegraphics[width=13cm]{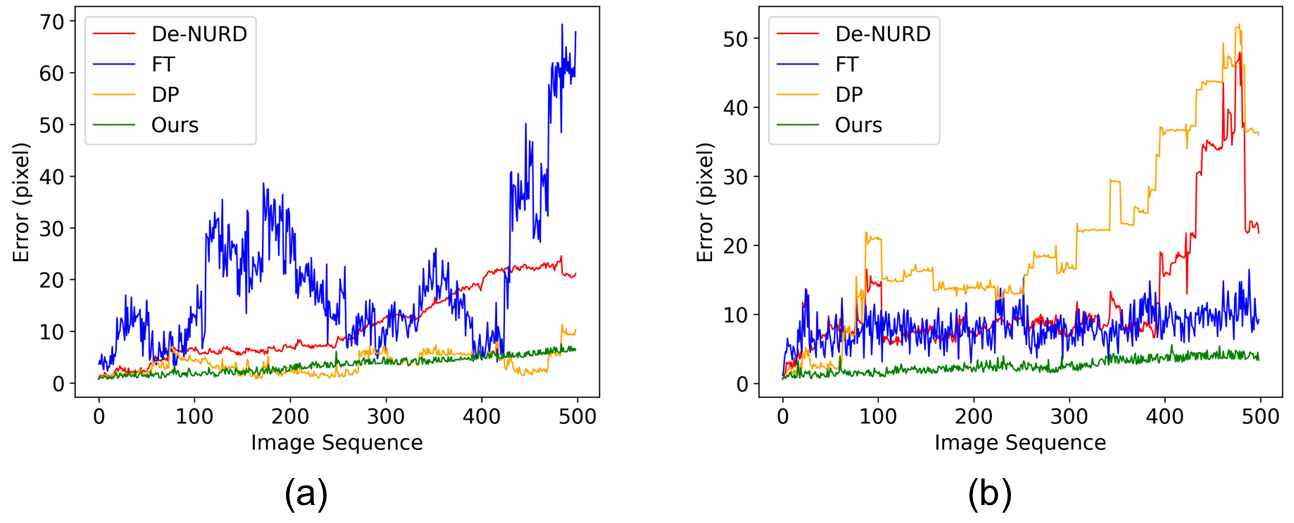}
    \caption{Quantitative comparison of different NURD correction methods using the two synthetic sequences. (a) is the result of the pig bronchus data. (b) is the result of the human nasopharynx data.}
    \label{fig:supplement1}
\end{figure}

\begin{figure}[H]
    \centering
    \centering\includegraphics[width=13cm]{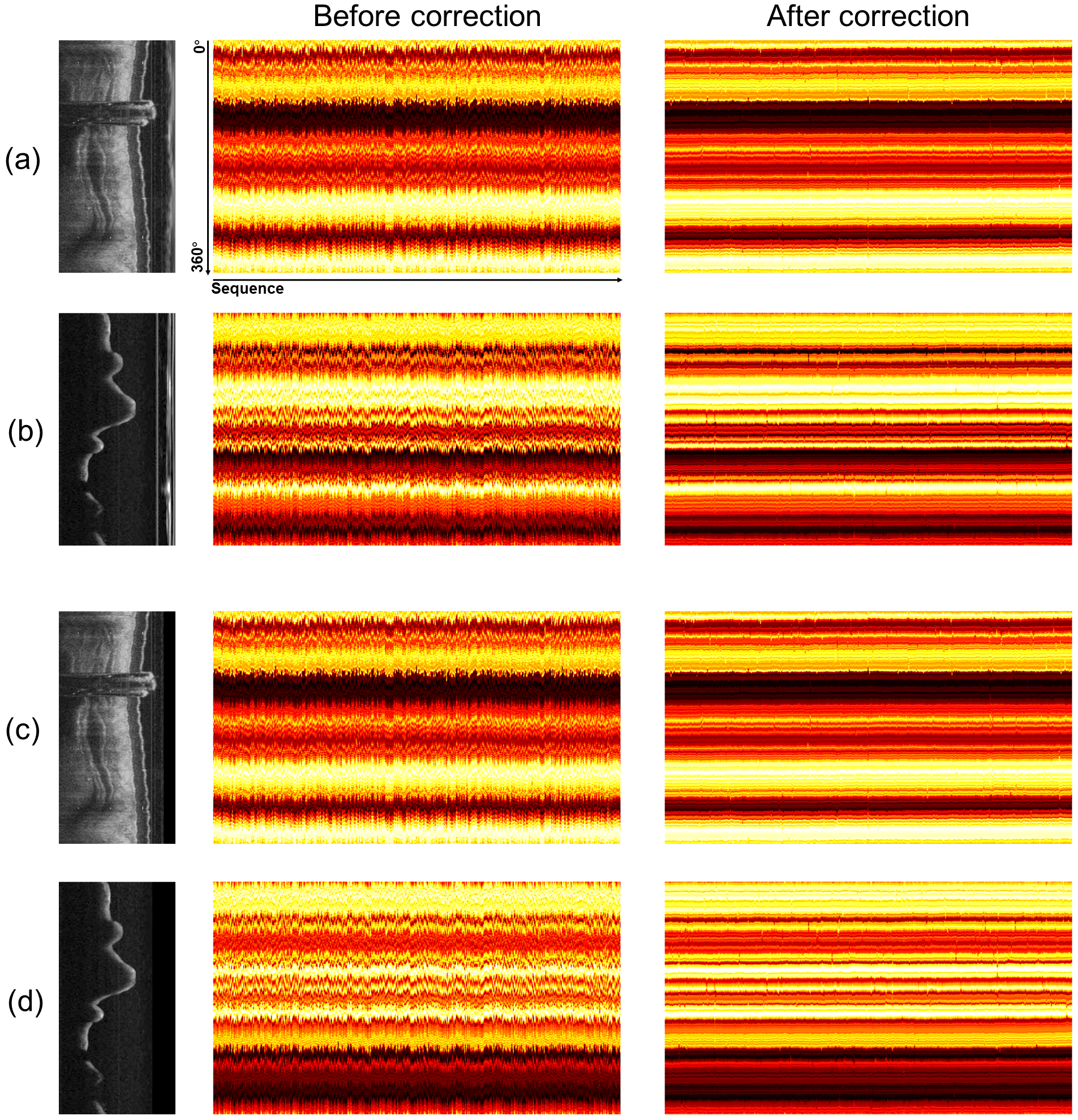}
    \caption{The NURD performance of two synthetic sequences. (a) is the result of the pig bronchus data. (b) is the result of the human nasopharynx data. (c) and (d) are the results after removing the tissue-unrelated features, such as sheath. The left column gives the original B-frames used to create the synthetic sequences. The middle column shows the axial maximum value projection of the synthetic sequences, which gives better views of the applied NURD. The right column gives the NURD-corrected synthetic sequences using our method.}
    \label{fig:supplement2}
\end{figure}

\indent Using the synthetic endoscopic OCT sequences that have the GT, we perform the quantitative comparison of our proposed method with three other representative approaches, including a feature tracking (FT) method \cite{lo2019balloon}, a dynamic programming (DP) method \cite{van2008azimuthal}, and the CNN-based method in \cite{liao2022distortion} (referred to as De-NURD). The results are shown in Table.~\ref{tab:my-table} and Fig.~\ref{fig:supplement1}. Our method achieves the smallest MAE values compared with the other three NURD correction methods on both synthetic sequences. Specifically, as shown in Fig.~\ref{fig:supplement1} (a) and (b), our method corrects the NURD with high accuracy and superior correction stability across the frames in each sequence. Other methods, in contrast, lack either correction accuracy or stability.
\begin{table}[H]
\centering
\caption{Quantitative comparison of different NURD correction methods using the two synthetic OCT sequences. The data format in the table is mean (standard deviation).}
\label{tab:my-table}
\begin{tabular}{ccc}
\hline
        & Pig bronchus           & Human nasopharynx      \\ \hline
De-NURD & 11.167 (6.799)         & 12.261 (9.277)         \\
FT      & 19.633 (13.921)        & 8.127 (2.352)          \\
DP      & 3.734 (2.038)          & 20.519 (12.537)        \\
Ours    & \textbf{3.489 (1.595)} & \textbf{2.561 (1.051)} \\ \hline
\end{tabular}
\end{table}
\indent Figure.~\ref{fig:supplement2} demonstrates the results before and after the NURD correction using our method, on (a) the pig bronchus data and (b) the human nasopharynx data. The left column gives the original B-frames used to create the synthetic sequences. The middle column shows the axial maximum value projection of the synthetic sequences, which gives better views of the applied NURD. The right column gives the NURD-corrected synthetic sequences using our method. As shown in the figure, our method alleviates the shift and jitter caused by the NURD while the original structure information is maintained. In addition, the NURD is effectively corrected on both the synthetic porcine bronchial sequence (a), which has rich feature information, and the human nasopharyngeal sequence (b), which has less feature information, suggesting that our method has superior robustness. To verify the NURD correction is performed on the features of biological tissues, we manually remove tissue-unrelated features (sheath, wire, etc.) in the data as shown in Fig.~\ref{fig:supplement2} (c) and (d). Under this condition, our method is still able to correct the NURD in both the human nasopharynx and the pig bronchus data.
\subsection{Robustness assessment of NURD correction}
\begin{figure}[H]
    \centering
    \centering\includegraphics[width=12cm]{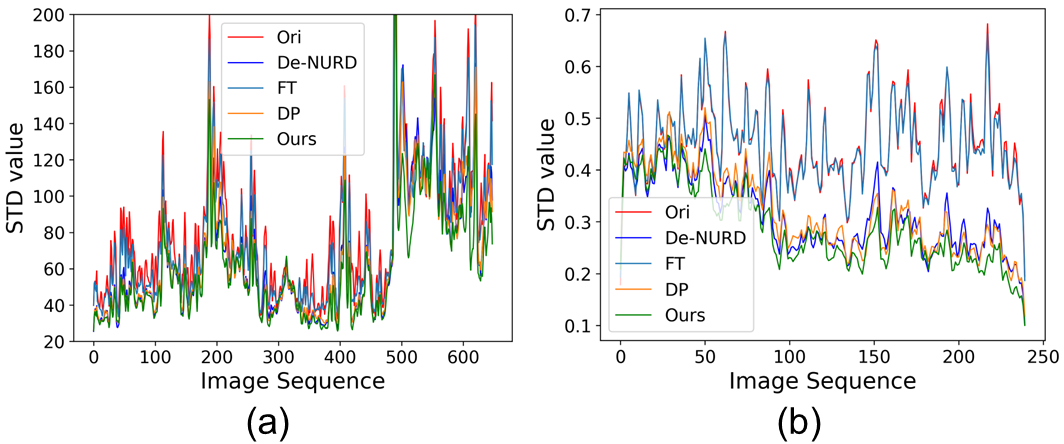}
    \caption{Quantitative comparison of different NURD correction methods using two publicly available datasets. (a) is the result of the gastrointestinal tract data. (b) is the result of the flat sponge surface data.}
    \label{fig:img4}
\end{figure}

\begin{figure}[H]
    \centering
    \centering\includegraphics[width=13cm]{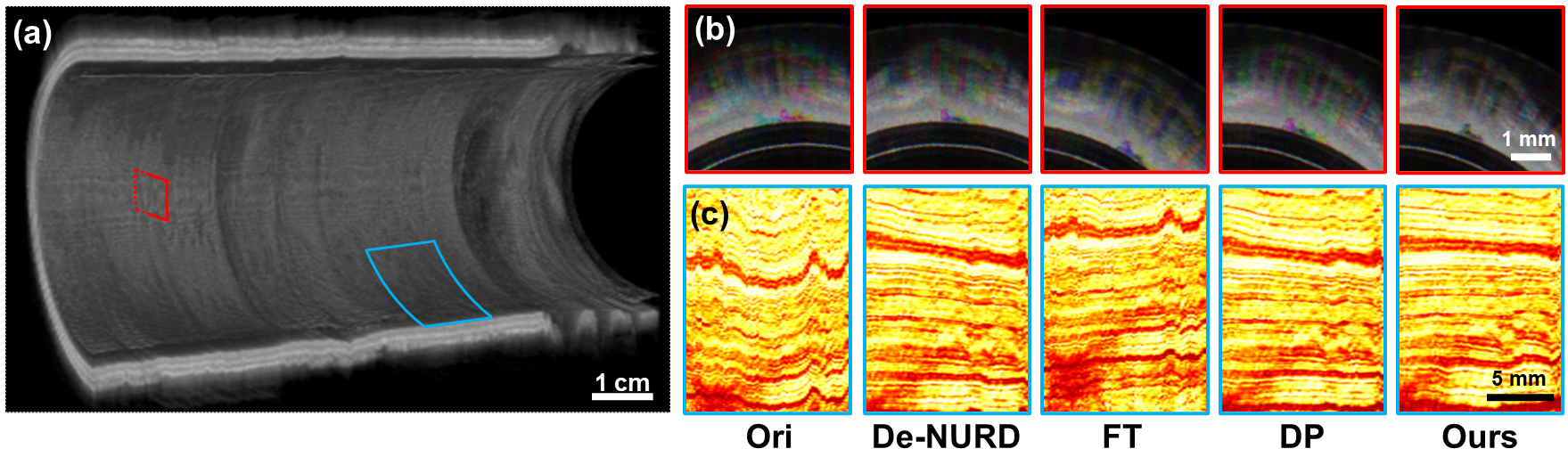}
    \caption{Qualitative comparison of different NURD correction methods on gastrointestinal tract test data. (a) is the 3D view of a volumetric scan of the gastrointestinal tract. The red and blue boxes refer to the zoom-in area in (b) and (c), respectively. (b) The local regions of OCT images are composed of three consecutive frames which are separately mapped to R, G, and B color channels. (c) Local \textit{en face} images with mean value projection operation.}
    \label{fig:img5}
\end{figure}

\begin{figure}[H]
    \centering
    \centering\includegraphics[width=8cm]{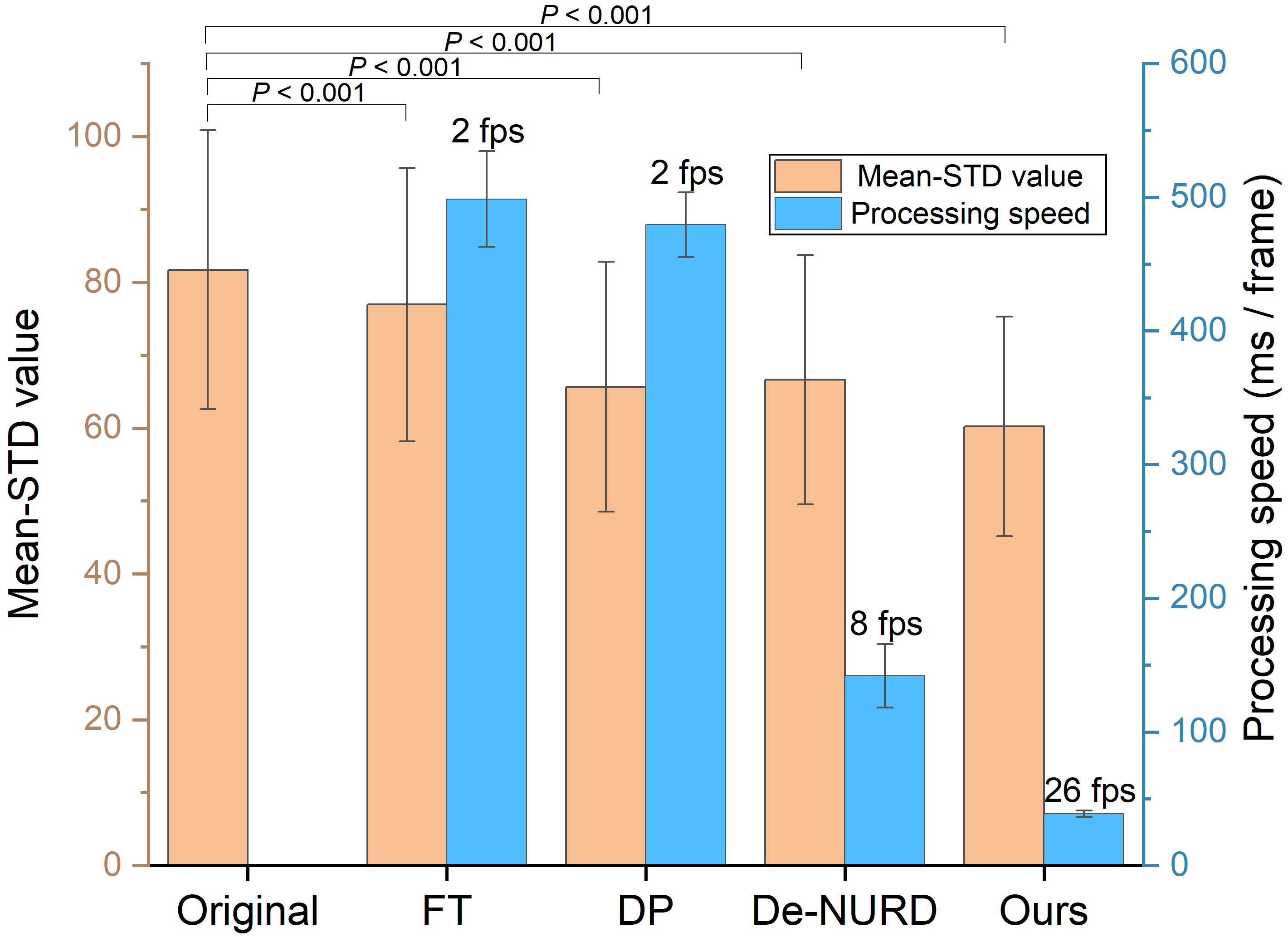}
    \caption{Comparison between the results of our proposed method and three other representative approaches. Orange bars represent their mean-STD (smaller means better). Blue bars represent the processing speed (ms$/$frame) and the corresponding frame rate in fps.}
    \label{fig:img7}
\end{figure}

 The results of the two real publicly-available testing datasets are shown in Table~\ref{tab:table1} and Figure~\ref{fig:img4}. Our proposed method achieves the smallest mean-STD values compared with the other three NURD correction methods \cite{lo2019balloon,van2008azimuthal,liao2022distortion}. Figure~\ref{fig:img4} (a) and (b) are the results of the gastrointestinal tract and the sponge surface data, respectively. The results of our method are plotted in green, demonstrating consistent minimum mean-STD values over the image sequences.
\begin{table}[H]
\centering
\caption{Quantitative comparison of different NURD correction methods using two publicly available datasets. The data format in the table is mean (standard deviation).}
\label{tab:table1}
\begin{tabular}{c|cc}
\hline
        & Gastrointestinal tract   & Sponge phantom         \\ \hline
Original     & 81.693 (38.261)          & 0.455 (0.081)          \\
De-NURD & 66.645 (34.217)          & 0.313 (0.072)          \\
FT      & 76.938 (37.481)          & 0.452 (0.079)          \\
DP      & 65.654 (34.302)          & 0.321 (0.082)          \\
Ours    & \textbf{60.225 (30.120)} & \textbf{0.288 (0.076)} \\ \hline
\end{tabular}%
\end{table}
\indent Figure~\ref{fig:img5} shows the qualitative comparison of different NURD correction methods on the gastrointestinal tract volume data. (a) is the 3D view of a volumetric scan of the gastrointestinal tract. The red and blue boxes refer to the zoom-in area in (b) and (c), respectively. In Fig.~\ref{fig:img5}(b), to illustrate the NURD instability, we use RGB channels to encode three consecutive frames, and each frame is mapped to an individual channel. Structures that do not overlap are rendered in color and vice versa in greyscale. We can see our method achieves the best spatial consistency. In Fig.~\ref{fig:img5}(c), we use mean value projection to obtain local \textit{en face} images. It can be seen that our method minimizes the distortion caused by the NURD.
\begin{figure}[b]
    \centering
    \centering\includegraphics[width=13cm]{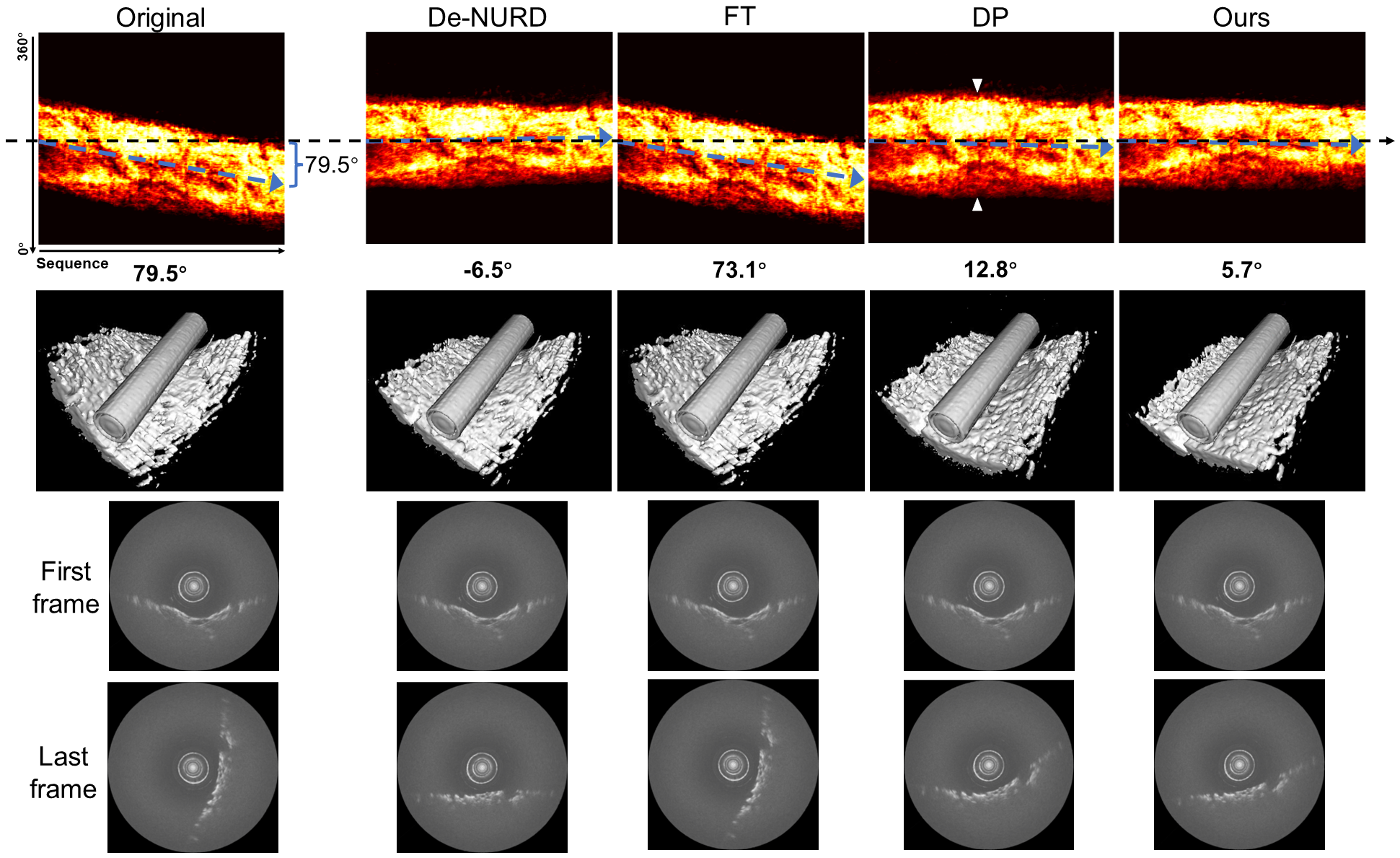}
    \caption{Qualitative evaluation of the NURD correction performance on a flat sponge surface data. \textit{En face} images by the mean value projection of the original and corrected results are shown in the first row, and the numbers at their bottom represent the NURD-induced precession angle of the flat surface. Their 3D rendering is shown in the second row. The last two rows are the first and last frames of the sequence.}
    \label{fig:img6}
\end{figure}
\indent Figure~\ref{fig:img6} presents the qualitative results of different NURD correction methods on pull-back scans of a flat surface of a sponge. The \textit{en face} images by the mean value projection of the original and corrected results are shown in the first row, and the numbers at their bottom represent the NURD-induced precession angle of the flat surface. It is obtained by (1) firstly connecting the center positions of the first and last frames in the sponge sequence (blue dashed line with arrow) and (2) then measuring the deviation angle between the blue dashed line and the flat reference (black dashed line). The original sequence is gradually distorted by NURD of synchronous rotation and pull-back scanning causing a maximum precession angle of 79.5$^\circ$. All the NURD correction methods are able to reduce the precession angle (a precession angle of $0^\circ$ represents the real state of the sponge). Our method outperforms others and achieves the minimum precession angle of $5.7^\circ$. Our method also reserves the morphological feature of the sponge, while the DP method causes structural stretch as pointed out by the white arrows. The second row of Fig.~\ref{fig:img6} is the 3D rendering of the sponge, which further illustrates the performance advantages of our method. Besides, we give the first and last frames of the sequence in the last two rows.\\
\indent To provide a comprehensive comparison of processing speed and correction accuracy, we combine the two and plot a histogram of the results of our proposed method against three other representative methods. The public gastrointestinal tract data is used in this evaluation. The results are shown in Fig.~\ref{fig:img7}. Orange bars represent their mean-STD (smaller means better). Blue bars represent the processing speed (ms/frame) and the corresponding frame rate in fps. It can be observed that our method achieves the best correction performance (statistically significant) while also improving processing speed by about three times, reaching real-time performance.
\subsection{Correction performance on 3D stent imaging}
\begin{figure}[H]
    \centering
    \centering\includegraphics[width=12cm]{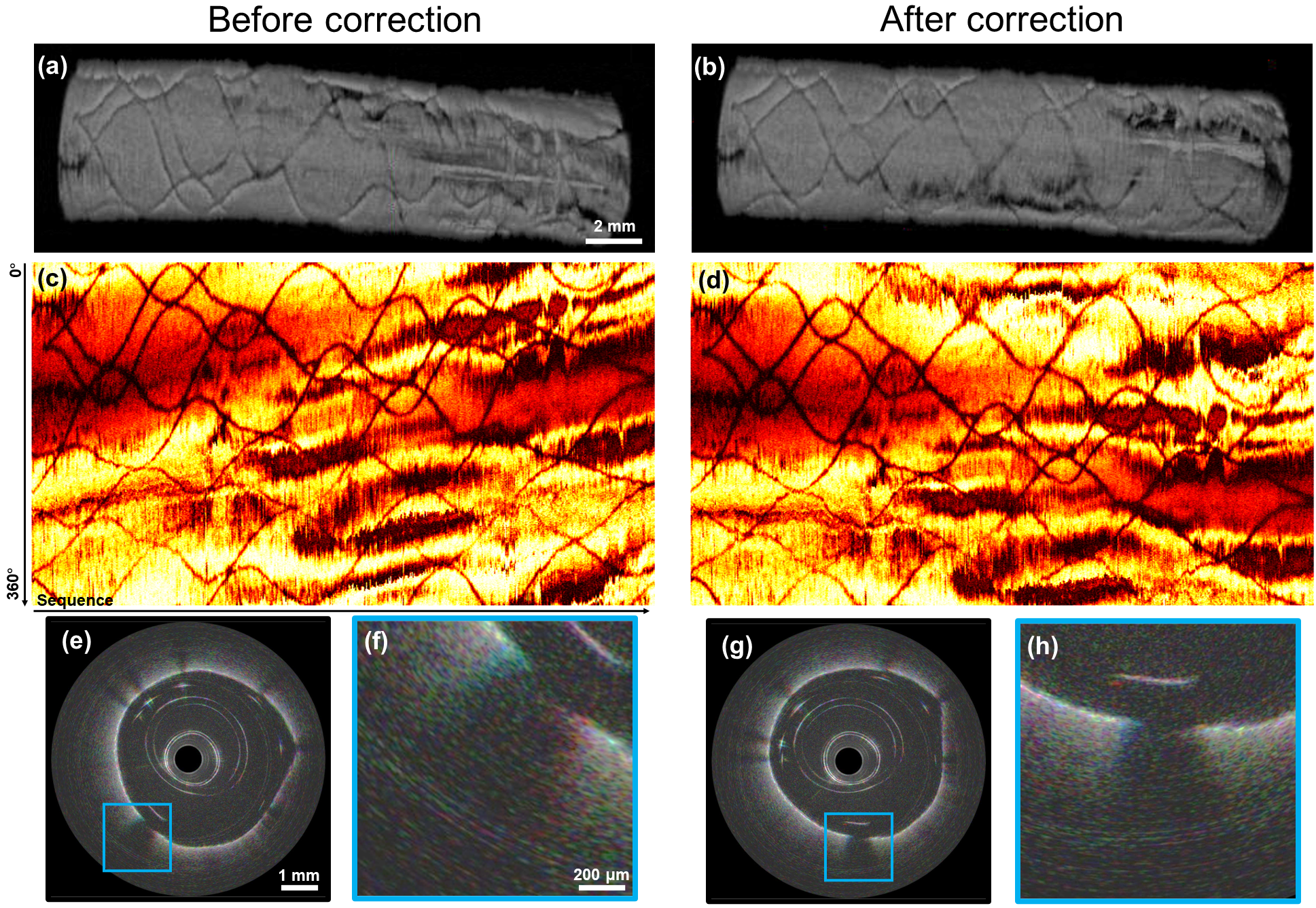}
    \caption{The NURD correction performance of 3D intravascular stent imaging. (a) and (b) are 3D views of the original data and corrected data, respectively. (c) and (d) are 2D \textit{en face} projections of (a) and (b). (e) and (g) are original and corrected cross-section images composed of three consecutive frames separately mapped in R, G, and B channels, respectively. (f) and (h) are enlarged views of the blue box in (e) and (g), respectively.}
    \label{fig:img8}
\end{figure}

\indent As a practical application, we conduct a pull-back endoscopic OCT scanning of the intravascular stent and correct distortion for the raw sequence to verify our correction performance for inherent NURD of the endoscopic OCT system. In vascular interventional procedures, endoscopic OCT imaging is commonly used to produce high-resolution in vivo images of blood vessels and deployed stents, providing accurate measurements of luminal architecture and insights regarding stent apposition \cite{ali2023optical,ali2016optical}. In this experiment, a 30 mm length intravascular stent with a 4 mm diameter was used for imaging shown in Fig.~\ref{fig:img3} (c). We wrapped up the stent with printer paper to simulate a lumen. We pulled back the mini-probe at a speed of 1.5 mm/s and collected about 640 images with $\sim$25 mm axial length. \\
\indent Figure~\ref{fig:img8} (a) and (b) show the 3D view of direct imaging and after correction, respectively. For an intuitive comparison, we unfold the 3D view to 2D \textit{en face} maps shown in Fig.~\ref{fig:img8} (c) and (d) by mean value projection. Due to friction and speed of the motor, shift distortion and uncertain stretch-shrink occur in the original \textit{en face} projection according to the inherent structure of the stent. After correction by our proposed method, The imaging appearance of the stent is closer to the real structure itself. In addition, we show a cross-section example of (e) before and (g) after correction at the same frame location with three consecutive frames mapped in 3 channels separately. The corresponding enlarged views are displayed in (f) and (h), respectively. By this, it can be observed that the proposed method alleviates the artifacts caused by NURD.
\subsection{Influence of training data}
\begin{figure}[b!]
    \centering
    \centering\includegraphics[width=13cm]{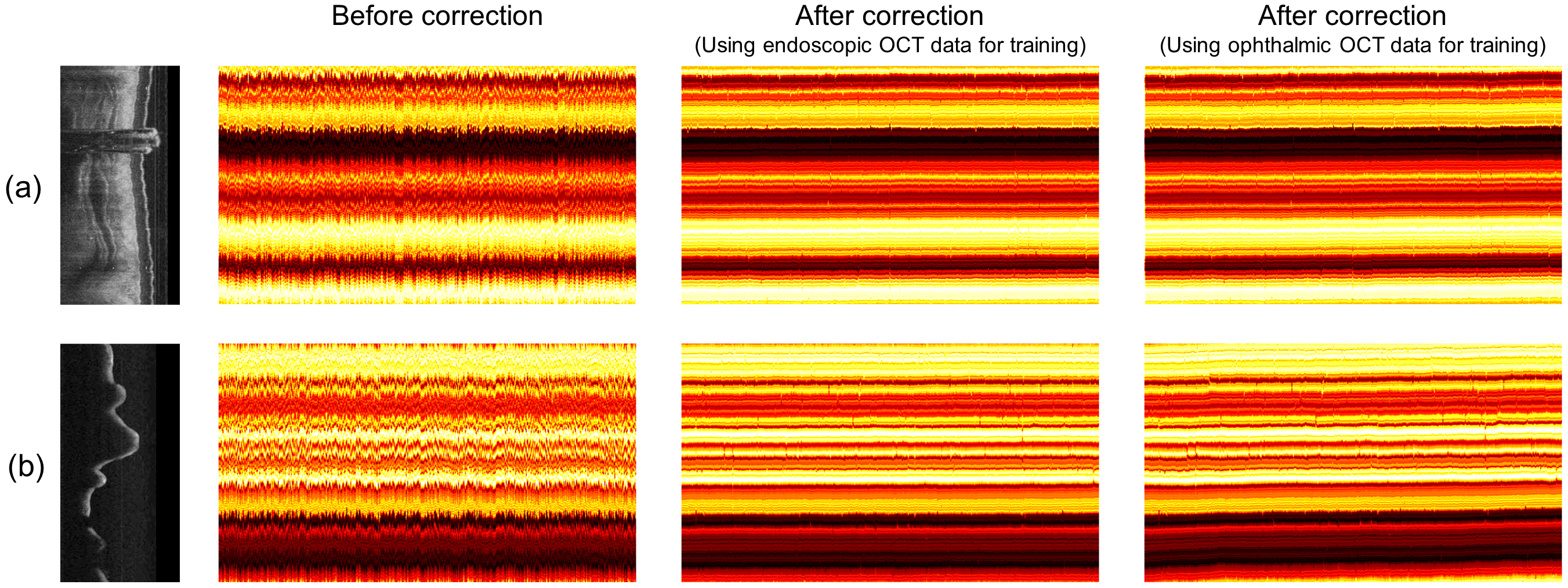}
    \caption{Qualitative comparison of NURD correction performance using endoscopic and ophthalmic OCT data for training. (a) The results of synthetic pig bronchus data. (b) The results of human nasopharynx data.}
    \label{fig:supplement4}
\end{figure}
\begin{figure}[H]
    \centering
    \includegraphics[width=\linewidth]{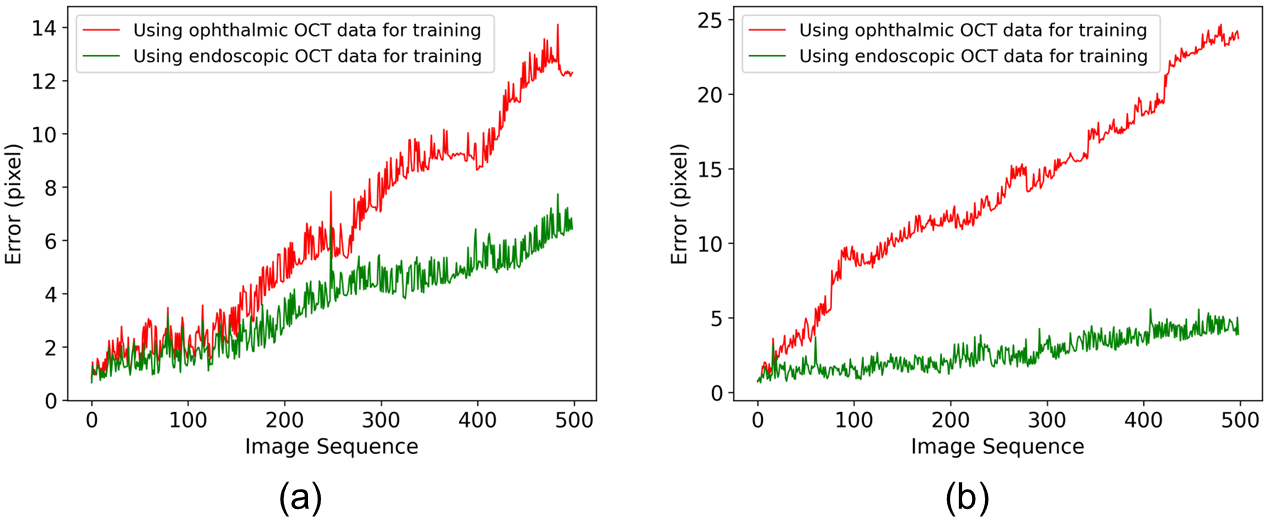}
    \caption{Quantitative comparison of NURD correction performance using endoscopic and ophthalmic OCT data for training. (a) The results of synthetic pig bronchus data. (b) The results of human nasopharynx data.}
    \label{fig:supplement3}
\end{figure}

\indent In the training of our NURD correction model, we use publicly-available endoscopic OCT datasets, which may influence the correction performance due to their intrinsic NURD. To address this issue, we also employ ophthalmic OCT data for training, which was acquired via raster scanning and thus inherently NURD-free. We employ 11,206 retinal OCT B-scans from a publicly-available dataset \cite{bogunovic2019retouch} with the same distortion vectors extracted from endoscopic OCT data to generate 20,000 original-distorted training pairs.\\
\indent The NURD correction results using these two types of training data are demonstrated in Fig.~\ref{fig:supplement4}, Fig.\ref{fig:supplement3}, and Table~\ref{tab:raster-scan test}. As shown in Fig.~\ref{fig:supplement4}, the NURD on the synthetic human nasopharynx and pig bronchus data could be effectively corrected when both the endoscopic and ophthalmic OCT data are used for training. The corresponding quantitative results (Fig.~\ref{fig:supplement3}) indicate better performances are achieved when using the endoscopic OCT data for training. We further deploy the trained models on the real gastrointestinal tract data. Their quantitative results are listed in Table.~\ref{tab:raster-scan test}. Consistent with the results of the synthetic data, the model trained on endoscopic OCT data performs better.
\begin{table}[H]
\centering
\caption{Comparison of the NURD correction on the real gastrointestinal tract data using different types of training data. The data format in the table is the mean (standard deviation).}
\label{tab:raster-scan test}
\begin{tabular}{ccc}
\hline
Original             & \begin{tabular}[c]{@{}c@{}}Using endoscopic OCT data\\for training\end{tabular} & \begin{tabular}[c]{@{}c@{}}Using ophthalmic OCT data\\for training\end{tabular} \\ \hline
81.693 (38.261) & \textbf{60.225 (30.120)}                                                                 & 64.585 (35.064)                                                             \\ \hline
\end{tabular}
\end{table}
\indent Due to the domain discrepancy between the endoscopic and ophthalmic OCT data, we can only suspect that the influence of the inherent NURD is neglectable. From the perspective of model training, our method (as illustrated in Fig.~\ref{fig:img1}) aims to align the designedly distorted frame with the original one. Whether or not the original frame is inherently distorted, the model is to predict the artificially created distortion vectors, the influence of the inherent NURD should be minimal.
\subsection{Ablation studies}
\indent To evaluate the effectiveness of the bi-directional prediction loss of two distortion vectors between the original frame and distorted frame in the training phase designed in our proposed method, we perform ablation studies on the gastrointestinal tract test data. \\
\indent The results evaluated with mean-STD metric are shown in Table~\ref{tab:table2}. For predicting the distortion vector 1 that transformed the original frame into the distorted frame, using only L1 Loss can significantly improve the performance of correcting distortion reducing mean-STD value of $\sim$19. When combined with the smoothness loss and similarity loss, the mean-STD value is slightly reduced. With the addition of another auxiliary prediction of distortion vector 2 that transformed the distorted frame into the original frame, the results achieve the best performance compared with other settings. It is noted that further improvements demonstrate the effectiveness of bi-directional prediction. Furthermore, this setting alleviates the NURD in the inference phase without adding much computation time and additional label burden in the training phase.
\begin{table}[H]
\centering
\caption{Ablation studies of different prediction loss settings. The data format in the table is the mean (standard deviation).}
\label{tab:table2}
\resizebox{\columnwidth}{!}{%
\begin{tabular}{cccc}
\hline
Original             & $\mathcal{L}_{l1}$              & $\mathcal{L}_{l1}+\mathcal{L}_{sm-1}+\mathcal{L}_{si-1}$          & $\mathcal{L}_{l1}+\mathcal{L}_{sm-1}+\mathcal{L}_{si-1}+\mathcal{L}_{sm-2}+\mathcal{L}_{si-2}$       \\ \hline
81.693 (38.261) & 62.420 (32.145) & 62.056 (31.178) & \textbf{60.225 (30.120)} \\ \hline
\end{tabular}%
}
\end{table}
\subsection{Evaluation of processing speed}
\indent Finally, we compare the processing speed of the two learning-based approaches in further detail. As shown in Table~\ref{tab:table3}, our method enables significant time-savings during pre/post-processing compared with the CNN-based method, which is due to the fact that our approach does not require the pre-construction of a spatial correlation matrix. Note that we achieve the capability of real-time NURD correction at 26$\pm$3 fps while keeping a good accuracy.
\begin{table}[H]
\centering
\caption{Comparison of the processing speed using two learning-based methods.}
\label{tab:table3}
\resizebox{\columnwidth}{!}{%
\begin{tabular}{cl|cccc}
\hline
\multicolumn{2}{c|}{Methods} & Pre- \& post-processing & Model inference      & Total time/frame       & Frame per second  \\ \hline
\multicolumn{2}{c|}{De-NURD} & 133.47±22.64 ms         & \textbf{3.11±1.68ms} & 136.28±23.81 ms        & 8±1 fps           \\ \hline
\multicolumn{2}{c|}{Ours}    & \textbf{29.81±3.63 ms}  & 8.86±0.48ms          & \textbf{38.67±3.64 ms} & \textbf{26±3 fps} \\ \hline
\end{tabular}%
}
\end{table}
\section{Discussion}
Self-attention, a groundbreaking mechanism for deep learning, has ushered in transformative advancements in NLP and CV \cite{niu2021review}. In NLP, large language models like BERT and GPT-4, built on self-attention, have excelled in language tasks due to their ability to capture context and dependencies in text \cite{patwardhan2023transformers}. In CV, the vision transformer architecture and its variants leverage self-attention to process images by dividing them into patches and applying this mechanism to them \cite{dosovitskiy2020image}. They have achieved remarkable success in many tasks, such as classification, object detection, and semantic segmentation \cite{han2022survey}. Besides, downstream applications such as medical image analysis also benefit from the paradigm shift from CNN to transformer \cite{li2023transforming}. \\
\indent In this work, we employ the self-attention mechanism to address the NURD problem in endoscopic OCT. We found that its capability of learning long-range dependencies and spatial correlations is useful in improving the efficiency of NURD correction. We designed the stacked cross-attention network specifically for this application (described in Section 2.2). Compared with existing NURD methods, our method achieves a $\sim3\times$ speedup to real time ($26\pm 3$ fps). We further design an overall framework for learning the NURD correction (described in Section 2.1) by leveraging three types of optimization constraints, including the L1, smoothness, and similarity losses. We also introduce a bi-directional design in the architecture of the framework. Their effectiveness in improving the NURD correction performance is verified through the ablation studies in Section 3.3. These new designs allow our method to outperform existing NURD correction methods not only in terms of efficiency but also in terms of performance.\\
\indent To verify the generalization performance of our method, we test it on the data from several different OCT systems that cover the mainstream engines for endoscopic OCT imaging, including: (1) A tethered capsule endomicroscope for imaging gastrointestinal tract using a swept-source OCT system \cite{gora2013tethered}. It uses near-infrared wavelengths sweeping from 1,250 nm to 1,380 nm. It acquires circumferential, cross-sectional images at 20 frames s$^{-1}$ using a total of 2,048 axial (depth) scans per image. (2) A volumetric scanning OCT system for general luminal organ diagnosis \cite{liao2021data}. It was built around the Axsun swept-source engine, with a 1310 nm center wavelength-swept source laser and 100 kHz A-line rate. The OCT probe has an outer diameter of 3.5 mm. It is terminated at the distal end with a transparent sheath on the tip, which allows three-dimensional OCT imaging using an internal rotating side-focusing optical probe with two proximal external scanning actuators. (3) A home-built endoscopic OCT system for intravascular imaging, which uses a spectral-domain OCT system for collecting the interference fringe. It has a central wavelength of 840 nm and a line rate of 80 kHz. The fiber-optic probe has an outer diameter of 0.46 mm. A homemade capillary tube-based fiber optic rotary joint driven by a commercial motor (34 rps rotation speed) is applied to perform circumferential scan imaging. For the data from the above systems, our method achieves superior accuracy and efficiency in the NURD correction.\\
\indent Our method can be beneficial to many application scenarios of OCT: (1) Surgical navigation and surveillance using OCT have revolutionized the field of minimally invasive procedures \cite{zaffino2020review,de2021assessment,yunyao2023review}. With its high-resolution imaging capabilities, OCT allows surgeons to navigate with unprecedented precision within complex anatomical structures. During surgery, real-time OCT imaging provides dynamic feedback, enabling surgeons to visualize tissue layers, assess boundaries, and confirm instrument placement. This real-time guidance enhances surgical accuracy, reduces the risk of complications, and minimizes the need for extensive tissue dissection. (2) Functional OCT imaging techniques that require capturing temporal dynamics (repeated scanning of a specific position), such as angiography, elastography, and thermometry. Bouma \textit{et al.} developed a microscopic image guidance platform for radiofrequency ablation (RFA) using a clinical balloon-catheter-based optical coherence tomography (OCT) system \cite{lo2019balloon}. They have shown that the computational correction of NURD could be used to improve the calculation of complex differential variance, which was then used to visualize the therapeutic thermal field. (3) The high spatial resolution of OCT enables its applications in rapid \textit{In situ} diagnosis. The presence of NURD increases the probability of misdiagnosis. Especially now that AI diagnostic models have been integrated with imaging instruments, the impact of imaging distortions will be further amplified \cite{leitgeb2021enhanced}.\\
\indent Despite the above merits, the NURD correction method based on the proposed cross-attention learning has some limitations: (1) Learning-based methods require a large number of labels (supervision) for training. As mentioned above, we follow the approach in \cite{liao2022distortion} by extracting the pseudo-GT distortion vectors using a feature-tracking method. Then we apply these distortion vectors randomly to the OCT images used in training. The stacked cross-attention network is trained to learn the mapping from manually distorted images to distortion-free ones. However, such a method is data-hungry and time-consuming. To address this issue, different supervision generation methods should be developed. (2) Our method is still in the category of image-based NURD correction and thus has the inherent drawbacks of such methods. This type of approach assumes that adjacent frames show a high degree of morphological coherence, \textit{i.e.}, and rotational artifacts result in faster changes in appearance than structural changes inherent in the appearance of the tissue. This is usually feasible in general clinical endoscopic imaging, except in a few cases, such as structural mutations and microscopic lesions at tissue junctions.
\section{Conclusions}
Here we tried to address the efficiency issue of NURD correction in endoscopic OCT and its functional extensions. Inspired by the self-attention mechanisms, we have developed a cross-attention learning method, to establish spatial correlations between OCT A-lines efficiently. We have designed and implemented an end-to-end stacked cross-attention network with optimization constraints. Compared to existing methods, we have achieved a substantial $\sim3\times$ speedup to real-time processing ($26\pm 3$ frames per second) and superior NURD correction performance. Our approach will contribute to the further development of endoscopic OCT technology and its multi-organ, multi-functional, multi-clinical scenario applications, as well as other rotational scanning imaging techniques such as intravascular ultrasound.
\begin{backmatter}
\bmsection{Funding}
This work was supported by the National Natural Science Foundation of China under Grant No.~51890892 and 62105198.
\bmsection{Acknowledgments}
We would like to thank the Editors and the anonymous Reviewers for their time and effort in helping us improve this manuscript.
\bmsection{Disclosures}
The authors declare no conflicts of interest.
\bmsection{Data availability}
Data used for training and test underlying the results presented in this paper are available in Ref. \cite{wang2015heartbeat,uribe2015rotational,gora2013tethered,sun2012vivo,lee2011quantification,miao2021graph,li2012high,wang2013intravascular,yun2006comprehensive,liao2021data}.
\end{backmatter}
\bibliography{sample}

\end{document}